\documentclass[12pt]{iopart}

\usepackage{iopams}
\usepackage{graphicx}
\newcommand{\A}{\cal A}
\newtheorem{proposition}{Proposition}
\newtheorem{theorem}{Theorem}
\newtheorem{lemma}{Lemma}

\begin{document}

\title{Binary trees, coproducts, and integrable systems}

\author{B Erbe$^1$ and H J Schmidt$^2$}

\address{$^1$Department of Physics, University of Regensburg,
Regensburg, Germany} \ead{bjoern.erbe@physik.uni-regensburg.de}
\address{$^2$Department of Physics, University of Osnabrueck,
Osnabrueck, Germany} \ead{hschmidt@uos.de}
\begin{abstract}
We provide a unified framework for the treatment of special
integrable systems which we propose to call ``generalized mean
field systems". Thereby previous results on integrable classical
and quantum systems are generalized. Following Ballesteros and
Ragnisco, the framework consists of a unital algebra with
brackets, a Casimir element, and a coproduct which can be lifted
to higher tensor products. The coupling scheme of the iterated
tensor product is encoded in a binary tree. The theory is
exemplified by the case of a spin octahedron. The relation to other
generalizations of the coalgebra approach is discussed.

\end{abstract}

\vspace{2pc}
This is an author-created, un-copyedited version of an article accepted for publication in Journal of Physics A: Mathematical and Theoretical. IOP Publishing Ltd is not responsible for any errors or omissions in this version of the manuscript or any version derived from it. The definitive publisher authenticated version is available online at 10.1088/1751-8113/43/8/085215. 
\maketitle

\section{Introduction}\label{sec:I}
In classical mechanics  ``complete integrability" can be precisely
defined in terms of the Arnol'd-Liouville theorem \cite{A78}. The
corresponding generalization of this concept to quantum theory has
not yet been achieved. Nevertheless, there exists a rich
literature on integrable quantum systems under various headlines
such as Yang-Baxter equations \cite{Bax82}, algebraic Bethe ansatz
\cite{Fad95} and quantum groups \cite{Dri87}. Aside from this
mainstream of research there are different theories of integrable
systems which could be characterized as ``generalized mean field
systems" (GMFS) \cite{BR98}, \cite{SS09}. It is the aim of the
present article to provide a general framework
for the description of such systems.\\

The prototype of the systems in question is a spin system where the
spins are (Heisenberg-) coupled to each other  with equal strength.
It turns out that each spin will move exactly as if it would be
under the influence of a uniform magnetic field. This justifies the
above characterization as (generalized) ``mean field systems". The
first generalizing step would be to consider systems which consist
of uniformly coupled integrable subsystems. This property can be
recursively applied. The underlying sequence of partial uniform
couplings is most conveniently encoded in a binary tree, the leaves
of which correspond to the smallest constituents of the system, see
\cite{SS09}. For example, the uniform coupling of three pairs of
spins can be described by the tree of figure \ref{fig0} and gives
rise to an integrable spin octahedron,
see figure \ref{fig4}.\\

Another generalization into a different direction is based upon
the insight that at the core a GMFS consists of a unital algebra
$\A$ with a bracket $[,\}$ and a co-multiplication $\Delta$, which
can be lifted to tensor products of $\A$ and applied to a Casimir
(central element) $c\in\A$, see \cite{BR98}. One then considers
representations of $\A$ generated by certain commutation relations
where the bracket $[,\}$ will either be represented by a Poisson
bracket (classical case)
or by the commutator of matrices (quantum case).\\

In our paper we simplify and generalize the approaches of
\cite{BR98} and \cite{SS09}. Thereby the restriction to Heisenberg
spin systems in \cite{SS09} is abolished by encorporating the
coalgebra ansatz of \cite{BR98}. Vice versa, the theory of
\cite{BR98} will be reformulated by using the language of binary
trees, and generalized from ``homogeneous trees" to general ones.
We also found that the postulate in \cite{BR98}
of $\Delta$ being ``co-associative" is superfluous, but see Appendix A.
After the first publication of the coalgebra approach \cite{BR98}
various generalizations have been proposed,
see \cite{BMR02}, \cite{BHMR04}, \cite{BB08}, \cite{M09a} and \cite{M09b}.
We will comment on the relation of our approach to these generalization
at the appropriate places in the paper and in two appendices.
The obvious generalization of assuming several
Casimir elements instead of a single one will be neglected here.\\

The paper is organized as follows. In section \ref{sec:T} we
collect some definitions concerning binary trees which are needed
later. Section \ref{sec:C} is devoted to the algebraic
prerequisites including the coproduct $\Delta$ and its lift
$\Delta^T$ to higher tensor products given by a binary tree $T$.
In section \ref{sec:M} we apply these tools to the theory of
integrable systems and prove the main result, theorem $1$, which
is analogous to prop.~$1$  of \cite{BR98} and provides a number of
commuting observables which is in many examples sufficient to
guarantee complete integrability. In section \ref{sec:E} we
discuss the elementary example of a Heisenberg spin octahedron in
order to illustrate the application of the abstract theory. Some
remarks of the corresponding Gaudin spin system and on the
connection to other approaches follow. Two appendices on the
issues of superintegrability and the recent loop coproduct
approach close the paper.

\section{Trees\label{sec:T}}

\begin{figure}
\includegraphics[width=8cm]{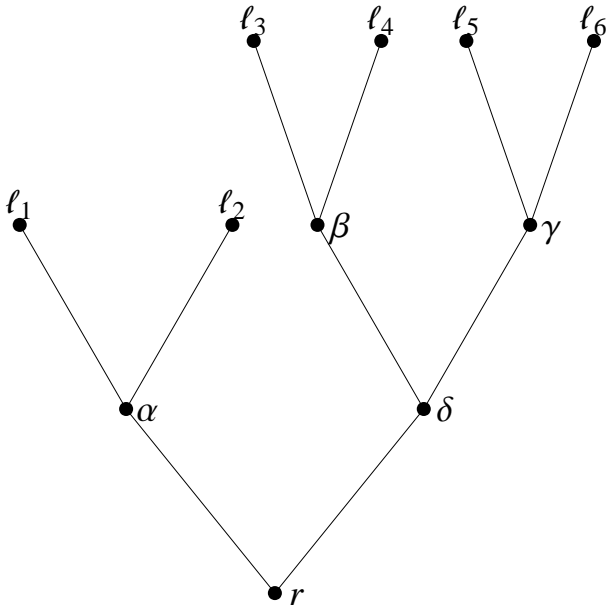}
\caption{\label{fig0}Example of a binary tree with root $r$, six
leaves $\ell_1,\ldots,\ell_6$ and further nodes
$\alpha,\beta,\gamma,\delta$.}
\end{figure}

We consider finite, binary trees $T$, shortly called ``trees".
Recall that these consist of a set of ``nodes" $\mathcal{N }(T)$,
such that all nodes $n\in\mathcal{N }(T)$, except the ``leaves"
$\ell\in\mathcal{L}(T)$, are connected to exactly two ``children"
$c_1(n),c_2(n)$, mixing the metaphors of horticulture and
genealogy. We have to distinguish between the ``left child" $c_1$
and the ``right child" $c_2$. Due to this distinction, the leaves
of a binary tree can be arranged in a natural order from left to
right and hence be labelled by ``$\ell_1$" to ``$\ell_L$". All
nodes, except the ``root" $r(T)$, are children of other nodes. By
definition, different nodes have different children, see figure
\ref{fig0}. As a tree, $T$ is a connected graph without cycles.\\

\begin{figure}
\includegraphics[width=8cm]{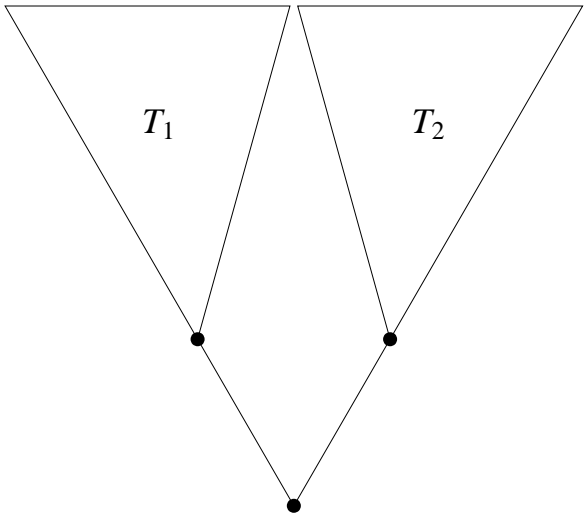}
\caption{\label{fig1}Union ${\sf V}(T_1,T_2)$ of two binary trees
$T_1$ and $T_2$.}
\end{figure}

The simplest tree $\bullet$ consists of only one root. The next
simplest one ${\sf V}$ has three nodes, that is, one root and two
leaves. If $T_1$ and $T_2$  are (disjoint) trees, then ${\sf
V}(T_1,T_2)$ will be the tree obtained by identifying the leaves
of ${\sf V}$ with the roots $r(T_1)$ and $r(T_2)$, see figure
\ref{fig1}.

\begin{figure}
\includegraphics[width=8cm]{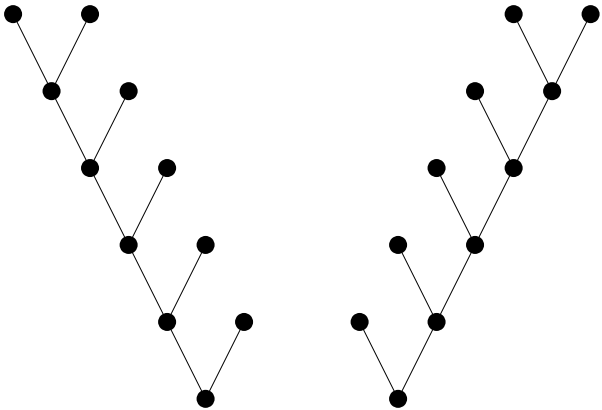}
\caption{\label{fig2}Examples of  ``homogeneous" binary trees.}
\end{figure}

Obviously, each tree can be obtained from copies of $\bullet$ by
recursively applying the operation  ${\sf V}(T_1,T_2)$. This opens
the possibility to provide recursive  definitions and proofs in
the theory of trees. A tree $T$ will be called ``homogeneous" if
it is of the form
\begin{eqnarray}\label{T1a}
T&=&{\sf V}(\ldots({\sf V}({\sf V}(\bullet,\bullet),\bullet),\ldots,\bullet)\\ \label{T1b}
\mbox{or } T&=& {\sf V}(\bullet,{\sf V}(\bullet,\ldots ,V(\bullet,\bullet)\ldots)
\;,
\end{eqnarray}
see, for example, figure \ref{fig2}. The tree of figure
\ref{fig0} is not homogeneous.\\

Binary trees are used in various parts of physics, e.~g.~in chaos
theory \cite{KK09}, computational physics \cite{WP02}, or in the
theory of spin networks \cite{TG09}. Here we utilize these
structures for encoding the
coupling schemes of certain integrable spin systems,
similarly as in \cite{FPJ94}, \cite{RJ09} and \cite{SS09}.\\

\begin{figure}
\includegraphics[width=8cm]{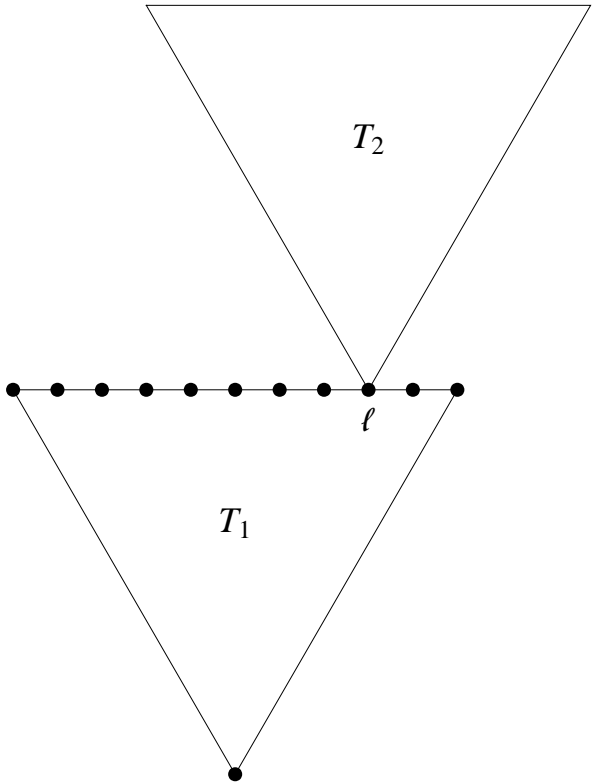}
\caption{\label{fig3}Binary tree $T=T_1\circ_\ell T_2$  obtained
by grafting $T_2$ on $T_1$.}
\end{figure}

The following lemma can be easily proved.
\begin{lemma}\label{L1}
$N(T)\equiv|\mathcal{N}(T)|=2|\mathcal{L}(T)|-1\equiv 2L(T)-1\;.$
\end{lemma}
A sub-tree $S\subset T$ is given by a subset of nodes of $T$,
which, according to their connections inherited from $T$ again
form a tree. For example, if $n\in \mathcal{N}(T)$ then $T(n)$
will denote the maximal sub-tree of $T$ with the root $n$. Let
$\mathcal{L}(n)\equiv \mathcal{L}(T(n))$. If
$n,m\in\mathcal{N}(T)$, then either
$\mathcal{L}(n)\subset\mathcal{L}(m)$ or
$\mathcal{L}(m)\subset\mathcal{L}(n)$ , or
$\mathcal{L}(n)\cap\mathcal{L}(m)=\emptyset$ .
In the former two cases $m$ and $n$ will be called ``connected", in the latter case ``disjoint".\\
If $T_1$ and $T_2$ are (disjoint) trees and
$\ell\in\mathcal{L}(T_1)$, then $T=T_1\circ_\ell T_2$ will denote
the tree obtained by ``grafting", i.~e.~by identifying the root
$r(T_2)$ with the leave $\ell$ of $T_1$, see figure \ref{fig3}.

\section{Coproducts\label{sec:C}}

In this paper we often will consider the classical and the quantum
case simultaneously. In both cases, the physical observables are
obtained by suitable representations of an abstract unital algebra
$(\mathcal{A},e)$ and its tensor products. In the quantum case,
$\mathcal{A}$ will be an associative, non-abelian algebra with
commutator $[a,b]=ab-ba,\; a,b\in\mathcal{A}$, and its physical
representation is given in terms of finite-dimensional matrices.
Typical examples are  cases where $\mathcal{A}$ is defined as the
universal enveloping algebra of some semi-simple Lie algebra. In
the classical case, $\mathcal{A}$ will be an abelian algebra
together with an abstract Poisson bracket $\{,\}$, see
\cite{BR98}. Representations of  $\mathcal{A}$ are then given by
the algebra
of smooth functions of some phase space together with the usual Poisson bracket.\\
To cover both cases, the commutator/Poisson bracket will be
denoted by $[a,b\},\; a,b\in\mathcal{A}$. It makes
$(\mathcal{A},[,\})$ into a Lie algebra and will act as an
derivation on the associative product on $\mathcal{A}$. We will
always consider algebras endowed with a bracket of one of these
two kinds and the corresponding homomorphisms, that is, linear
algebra
homomorphism w.~r.~t.~both multiplications.\\
If $\mathcal{A}_1, \mathcal{A}_2$ are two algebras as explained
above, then $\mathcal{A}_1\otimes \mathcal{A}_2$ will denote the
algebraic tensor product, physically describing a composite
system. It will be again a unital algebra with brackets upon
linearly extending the definitions
\begin{equation}\label{C1}
(a\otimes b)(c\otimes d)=(ac)\otimes(bd)
\end{equation}
and
\begin{equation}\label{C2}
[(a\otimes b),(c\otimes d)\}=[a,c\}\otimes\frac{bd+db}{2}+
\frac{ac+ca}{2}\otimes [b,d\} \;.
\end{equation}
If $A:\mathcal{A}_1\longrightarrow\mathcal{A}_1$ and
$B:\mathcal{A}_2\longrightarrow\mathcal{A}_2$  are morphisms as
explained above, then also $A\otimes B:\mathcal{A}_1\otimes
\mathcal{A}_2\longrightarrow\mathcal{A}_1\otimes \mathcal{A}_2$
will be such a morphism. \\
The ``coproduct" $\Delta$ will be a morphism
\begin{equation}\label{C3}
\Delta:\mathcal{A}\longrightarrow\mathcal{A}\otimes \mathcal{A}
\;,
\end{equation}
that is, a linear algebra morphism plus a Poisson bracket morphism
in the classical case. Usually, a coproduct is additionally
required to be ``co-associative", see \cite{BR98}, but this
property will not be needed in the main part of the present paper,
hence here we use the term ``coproduct" in a more general sense.
In Appendix A, co-associativity and co-commutativity of $\Delta$
will be assumed to extend the set of integrals of motion (superintegrability).\\

The crucial construction for integrability, as considered here, is
the lift of the coproduct to higher order tensor products given by
a tree $T$. To this end we first define $\mathcal{A}^T$
recursively by
\begin{eqnarray}\label{C4a}
\mathcal{A}^\bullet &=& \mathcal{A}\\ \label{C4b}
\mathcal{A}^{{\sf V}(T_1,T_2)} &=& \mathcal{A}^{T_1}\otimes
\mathcal{A}^{T_2} \;.
\end{eqnarray}
Sometimes it will be convenient to use the identification
\begin{eqnarray}\label{C5}
\mathcal{A}^{T_1} &=& \mathcal{A}^{T_2}
=\underbrace{\mathcal{A}\otimes\ldots\otimes\mathcal{A}}_{L\mbox{
times}} \;,
\end{eqnarray}
if $L(T_1)=L(T_2)\equiv L$. W.~r.~t.~this identification the canonical
embedding
\begin{equation}\label{C6}
j_n:\mathcal{A}^{T(n)}\longrightarrow \mathcal{A}^T,\quad
n\in\mathcal{N}(T)
\end{equation}
can be defined by
\begin{equation}\label{C7}
j_n(a)=e\otimes\ldots \otimes a\otimes\ldots \otimes e,\quad
a\in\mathcal{A}^{T(n)}\;.
\end{equation}

In the next step we define the lift of the coproduct
$\Delta^T:\mathcal{A}\longrightarrow\mathcal{A}^T$ recursively by
\begin{eqnarray}\label{C8a}
\Delta^\bullet&=&\mbox{id}_{\A} \\ \label{C8b} \Delta^{{\sf
V}(T_1,T_2)}&=& (\Delta^{T_1}\otimes\Delta^{T_2})\circ \Delta \;,
\end{eqnarray}
and conclude:
\begin{lemma}\label{L2}
$\Delta^T:\mathcal{A}\longrightarrow\mathcal{A}^T$ is a (Poisson)
algebra morphism.
\end{lemma}
\noindent{\em Proof.} By induction over $T$. The claim follows
since the tensor product and the composition of  (Poisson) algebra
morphisms is again a (Poisson) algebra morphism. \hfill$\Box$\\

Before formulating the main result we still need another
definition. Let $n\in\mathcal{N}(T)$, then
\begin{equation}\label{C9}
\Delta^n\equiv j_n\circ
\Delta^{T(n)}:\mathcal{A}\longrightarrow\mathcal{A}^T \;.
\end{equation}

We note that the generalization of the coalgebra approach to comodule
algebras \cite{BMR02} where $\Delta^n$ is replaced by a suitable map
${\mathcal A}\longrightarrow {\mathcal A}\otimes{\mathcal B}\otimes\ldots\otimes{\mathcal B}$
is only possible for homogeneous trees.\\

\section{Integrable systems\label{sec:M}}

Remember the grafting of trees $T=T_1\circ_\ell T_2$ explained in
section \ref{sec:T}. It gives rise to a corresponding composition
of lifted coproducts in the following sense:
\begin{lemma}\label{L3}
Let $T=T_1\circ_\ell T_2$, $x\in\mathcal{A}$ and write
$\Delta^{T_1}(x)=\sum_i x_{i,\ell_1}\otimes\ldots
x_{i,\ell_\mu}\otimes\ldots\otimes x_{i,\ell_{L_1}}$, where
$\ell_1,\ldots,\ell_{L_1}$ denote the leaves of $T_1$ and
$\ell_\mu=\ell$. Then $\Delta^T(x)=\sum_i
x_{i,\ell_1}\otimes\ldots
\Delta^{T_2}(x_{i,\ell_\mu})\otimes\ldots\otimes
x_{i,\ell_{L_1}}$.
\end{lemma}
\noindent{\em Proof.} By induction over $T_1$.
\hfill$\Box$\\

Further one can easily prove the following lemma:

\begin{lemma}\label{L4}
Let $n,m\in\mathcal{N}(T)$ be disjoint,
i.~e.~$\mathcal{L}(n)\cap\mathcal{L}(m)=\emptyset$ and
$x,y\in\mathcal{A}$. Then $[\Delta^n(x),\Delta^m(y)\}=0$.
\end{lemma}
\noindent{\em Proof.} $\Delta^n(x)$ and $\Delta^m(y)$ live in
disjoint factors of the tensor product $\mathcal{A}^T$, since they
are of the form $\Delta^n(x)=\sum_i
e\otimes\ldots\bigotimes_{\ell\in\mathcal{L}(n)} x_{i,\ell}
\otimes\ldots\otimes e$ and $\Delta^m(y) =\sum_j e \otimes \ldots
\bigotimes_{\ell\in\mathcal{L}(m)} y_{j,\ell} \otimes\ldots\otimes
e$ . Hence they commute.
\hfill$\Box$\\

Now we are ready to formulate the main result.
\begin{theorem}\label{M0}
Let $n,m\in\mathcal{N}(T)$ such that $\mathcal{L}(m)\subset\mathcal{L}(n)$,
$x\in\mathcal{A}$ and $c\in\mathcal{A}$ be a central element, i.e. $[y,c\}=0$ for all $y \in \mathcal{A}$. Then
$[\Delta^n(x),\Delta^m(c)\}=0$.
\end{theorem}

\noindent{\em Proof.} We
introduce the canonical partial embedding
$j_m^n:\mathcal{A}^{T(m)}\longrightarrow\mathcal{A}^{T(n)}$ such
that $j_m=j_n\circ j_m^n$. It follows that
\begin{eqnarray}
 [\Delta^n(x),\Delta^m(c)\} 
&=& [j_n\circ\Delta^{T(n)}(x),j_m\circ \Delta^{T(m)}(c)\}\\ \label{M1a}
&=&j_n\left( [\Delta^{T(n)}(x),j_m^n\circ \Delta^{T(m)}(c)\}
\right)\label{M1b}
\end{eqnarray}
and thus it suffices to show that
\begin{equation}\label{M2}
[\Delta^{T(n)}(x),j_m^n\circ \Delta^{T(m)}(c)\}=0 \;.
\end{equation}
By applying lemma \ref{L3} to the sub-tree $T(n)$ we write
\begin{equation}\label{M3}
T(n)=T_1 \circ_m T(m),\quad m=\ell_\mu\in\mathcal{L}(T_1)
\end{equation}
and conclude
\begin{eqnarray}
\Delta^{T(n)}(x)  
&=& \sum_i x_{i,1}\otimes\ldots
\Delta^{T(m)}(x_{i,\mu})\otimes\ldots x_{i,L_1}\;.\label{M4}
\end{eqnarray}
Hence
\begin{eqnarray}
[\Delta^{T(n)}(x),j_m^n\circ \Delta^{T(m)}(c)\} 
\nonumber &=&
\sum_i x_{i,1}\otimes\ldots
[\Delta^{T(m)}(x_{i,\mu}),\Delta^{T(m)}(c)\}\otimes\ldots
x_{i,L_1}\\ \nonumber
&=& \sum_i x_{i,1}\otimes\ldots \Delta^{T(m)}\left([x_{i,\mu},c\}\right)\otimes\ldots x_{i,L_1}\\
\label{M5} &=&0\;,
\end{eqnarray}
since $[x_{i,\mu},c\}=0$.
\hfill$\Box$\\

This theorem generalizes  prop.~$1$  of \cite{BR98} to arbitrary,
not necessarily homogeneous trees. In order to guarantee complete integrability
in the sense of the Arnol'd-Liouville theorem
for $2L$-dimensional phase spaces
(which is satisfied for spin systems, see section \ref{sec:E})
we would need $L$ pairwise commuting observables (``integrals in
involution"). These are provided by the $\Delta^n(c)$ for each
node $n$ which is not a leave. By lemma 1, there are exactly
$N(T)-L(T)=L(T)-1$ such nodes.
In \cite{BR98} the remaining observable is chosen as the Hamiltonian $H$.
In the context of quantum spin systems another choice would be more appropriate,
namely $\Delta^r(x)$ with a suitable $x\in{\mathcal A}$.
The Hamiltonian $H$ could then be chosen as any element of the algebra generated by the
$\Delta^n(c)$ and $\Delta^r(x)$.\\
In the general case the dimension of the phase space depends on the symplectic realization
of $({\mathcal A},\{,\})$ and the choice of the symplectic leaves. A thorough discussion
of these questions, which also applies to the binary tree approach,
including issues of superintegrability can be found in \cite{BB08}.
\\

\section{Examples and Outlook\label{sec:E}}

\begin{figure}
\includegraphics[width=8cm]{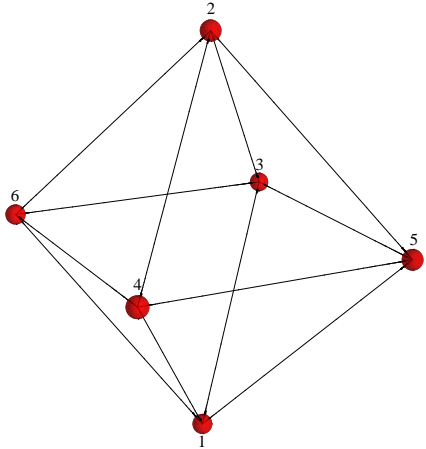}
\caption{\label{fig4}The octahedral spin graph corresponding to
the integrable Heisenberg Hamiltonian (\ref{M9}). Its coupling
scheme is encoded in the binary tree of figure \ref{fig0} as
explained in the text.}
\end{figure}

In order to  explain the application of theorem $1$ to
integrability of quantum systems we consider the elementary
example of a spin octahedron, figure \ref{fig4}, with Heisenberg
Hamiltonian, following \cite{SS09}. We chose as $\A$ the universal
enveloping algebra of the Lie algebra $su(2)$. More concretely: We
consider three generators $X_1,X_2,X_3$ satisfying the abstract
commutations relations
\begin{equation}\label{M6}
[X_j,X_k]=i \sum_{\ell=1}^3 \epsilon_{jk\ell} \, X_\ell \;,
\end{equation}
where $\epsilon_{jk\ell}$ denotes the completely anti-symmetric
Levi-Civita symbol. $\A$ is the set of all finite polynomials $X$
of the standard form
\begin{equation}\label{M7}
X=\sum_{klm}c_{klm}\,X_1^k\,X_2^l\,X_3^m \;.
\end{equation}
The product $XY$ of two such polynomials is brought into the
standard form (\ref{M7}) by successively applying the commutation
relations (\ref{M6}). The unit element in $\A$ is
$e=X_1^0\,X_2^0\,X_3^0$.
It follows that $c\equiv X_1^2+X_2^2+X_3^2$ commutes with all $X\in\A$.\\

The coproduct $\Delta$ is defined on the generators by
\begin{equation}\label{M8}
\Delta(X_i)=e\otimes X_i+X_i\otimes e
\end{equation}
and then extended to general elements of the form (\ref{M7}) by
employing the property of $\Delta$ being an algebra homomorphism.
Thus, for example,
\begin{eqnarray}
\Delta(c)&=& \Delta(X_1^2+X_2^2+X_3^2)\\
&=& \Delta(X_1)^2+\Delta(X_2)^2+\Delta(X_3)^2\\
&=& \sum_{i=1}^3 (e\otimes X_i+X_i\otimes e)^2\\
&=& e\otimes c + c\otimes e + 2\sum_{i=1}^3 X_i\otimes X_i \;.
\end{eqnarray}

We further choose $T$ as the binary tree of figure \ref{fig0} and
obtain the corresponding various commutations relations of theorem
$1$ being valid in the $6$-fold tensor product $\bigotimes_{i=1}^6
\A$ where one usually chooses $\Delta^r(x)=\Delta^r(X_3)$. Next we
consider the well-known $(2s+1)$-dimensional irreducible matrix
representation of (\ref{M6}) and denote the representations of the
generators $X_i$ by $\mathbf{S}_i$ (``spin operator components").
In the $6$-fold tensor product we denote the single spin
components by $\mathbf{S}_i^\mu,\,\mu=1,\ldots,6$. In this
representation, $c=s(s+1)\mathbf{1}$ and all commutation relations
of theorem $1$ remain valid. Note that $\Delta(c)$ becomes
$\left(\mathbf{S}^{(1)}+ \mathbf{S}^{(2)}\right)^2$, which is no
longer a constant, analogously for higher tensor products
$\Delta^n(c)$. This shows, by the way, why it is advantageous to
work in an abstract setting and to consider concrete
representations only after the coproduct is defined.  Note
further, that we could slightly generalize the example by
considering different $s$ for each factor
of the tensor product.\\

Let $(\mathcal{V},\mathcal{E})$ be the the octahedral spin graph
of figure \ref{fig4} with its set of  $6$ vertices $\mathcal{V}$
and the set of $12$ edges $\mathcal{E}$. The corresponding
Heisenberg Hamiltonian $H$ can be written in various ways:
\begin{eqnarray}\label{M9}
&&H= 2 J
\sum_{(\mu,\nu)\in\mathcal{E}}\mathbf{S}^{(\mu)}\cdot\mathbf{S}^{(\nu)}\\
\nonumber &&= J\left( \left(\sum_{\mu\in\mathcal{V}}
\mathbf{S}^{(\mu)}\right)^2
-\left(\mathbf{S}^{(1)}+\mathbf{S}^{(2)}\right)^2
-\left(\mathbf{S}^{(3)}+\mathbf{S}^{(4)}\right)^2
-\left(\mathbf{S}^{(5)}+\mathbf{S}^{(6)}\right)^2
\right)\\
&&\\
&&=J\left(
\Delta^r(c)-\Delta^\alpha(c)-\Delta^\beta(c)-\Delta^\gamma(c)
\right) \;,
\end{eqnarray}
where the root $r$ and the nodes $\alpha,\beta,\gamma$ refer to
the binary tree of figure \ref{fig0}. $J$ is some appropriate
coupling constant. It is crucial that $H$ can be written as a
linear combination of commuting observables according to theorem
$1$. In this respect the octahedral Heisenberg Hamiltonian
(\ref{M9}) is only the simplest case; for example, a Zeeman term
proportional to $\Delta^r(\mathbf{S}_3)$ could be added without
loosing integrability. The eigenvalues and common eigenvectors of
the system of commuting observables
$\Delta^n(c),\,n\in\mathcal{N}(T),$ result from the well-known
rules of coupling angular momenta involving Clebsch-Gordan
coefficients. An explicit formula for the eigenvalues and
eigenvectors of $H$ and arbitrary binary trees has been given in
\cite{SS09}. The example of the spin octahedron clearly shows the
physical meaning of the binary tree $T$ on which theorem $1$
depends: $T$ encodes the coupling scheme of systems which are
completely integrable
due to their structure of uniformly coupled subsystems.\\
With exactly the same algebraic considerations and the same tree
as above, theorem \ref{M0} provides us with another very
interesting integrable model:
\begin{eqnarray}
\label{Gaudin}
\nonumber H&=&A\left(\mathbf{S}^{(1)}\cdot \sum_{i=3}^6 \mathbf{S}^{(i)}
+ \mathbf{S}^{(2)}\cdot \sum_{i=3}^6 \mathbf{S}^{(i)}\right)
+ 2 (A+J)\mathbf{S}^{(1)} \cdot \mathbf{S}^{(2)} \\
&=& A\left(\Delta^r (c)-\Delta^{\delta}(c)\right)+J
\Delta^{\alpha}(c),
\end{eqnarray}
The (Gaudin) Hamiltonian $H$ describes a central spin system with
two central spins of exchange $2  (A+J)$, coupled homogenously to
a bath of four spins. Such a system can serve for example as a
simplified model for the hyperfine interaction in a
double quantum dot, see \cite{SKL03}.\\
Apart from this physical meaning, it is interesting from a formal
point of view. Besides the approach presented in this paper,
systems can be integrable in the sense of algebraic Bethe ansatz.
According to the ground breaking work of V.~Drinfel´d
\cite{Dri87}, this is based on quasicocommutative bialgebras,
which essentially means that there is an element $R \in \A \otimes
\A$ with
\begin{equation}
\left(\tau \circ \Delta\right)(x) \cdot R=R \cdot \Delta(x)
\end{equation}
for all $x\in \A$. $\tau$ denotes the switch operator
defined by linearly extending $\tau(a\otimes b)=b\otimes a$ and $\Delta$
a usual coproduct.
\newline As this algebraic structure is somewhat
similar to the one presented in this paper, the question arises,
whether there is a connection between systems integrable in either
sense. The above system, in contrast to the central spin system
with one central spin, is not integrable by means of algebraic
Bethe ansatz. Hence adding a second central spin destroys the Bethe
ansatz, whereas integrability in the sense of theorem \ref{M0}
remains unaffected.\\
Recently, a framework for integrability using so-called ``loop coproducts" has been
proposed \cite{M09a}, \cite{M09b} which contains different previous approaches to
integrability as special cases. It is, however, confined to the classical case. Some remarks on
the relation between this approach and the present paper are included in the appendix B.

\section*{Acknowledgements}
We are indebted to a referee for hints to the literature concerning generalizations of \cite{BR98}
and to Roman Schnalle for references on coupling trees. 

\appendix
\section{Superintegrability}

The approach \cite{BR98} to integrability via coalgebras has
subsequently be extended to ``superintegrability" \cite{BHMR04}, \cite{BB08}.
This roughly means that one is seeking for additional integrals of motion which, however,
do not longer commute with the old ones. ``Additional" means, in the classical case,
that the new integrals of motion are functionally independent of the old ones. Typically,
this functional independence cannot be shown in the general setting, but only in concrete
examples, see \cite{BHMR04}, \cite{BB08}. Also in theses references the role of co-associativity
of $\Delta$ in connection to superintegrability has been stressed.\\

The question arises, whether these ideas can be transferred to the more general situation
where the binary trees are not necessarily of homogeneous type. To this end we slightly refine
the binary tree construct, following, for example, \cite{FPJ94}. Recall that, due to the
distinction between ``left child" and ``right child", the leaves of a binary tree can be arranged
in a natural order from left to right and hence be labeled by ``$\ell_1$" to ``$\ell_L$".
Now assume that this labeling can be arbitrarily permuted. We will call the resulting structure
a \underline{coupling tree}. For example, the only binary tree with three nodes, {\sf V}, gives rise
to two different coupling trees, denoted by ${\sf V}(\ell_1,\ell_2)$ and ${\sf V}(\ell_2,\ell_1)$.
Generally, a coupling tree $\tilde{T}$ can be represented as a pair $\tilde{T}=(T,\pi)$, where $T$ is a binary
tree with $L$ leaves and $\pi\in{\mathcal S}_L$, a permutation of $L$ elements.
A coupling tree with $L$ leaves can alternatively be construed as a monomic expression in the
abstract variables $\ell_1,\ldots,\ell_L$, such that each variable occurs exactly once. For example,
the tree of figure \ref{fig0}, conceived as a coupling tree, corresponds to the expression
$(\ell_1\,\ell_2)((\ell_3\,\ell_4)(\ell_5\,\ell_6))$.\\

\begin{figure}
\includegraphics[width=10cm]{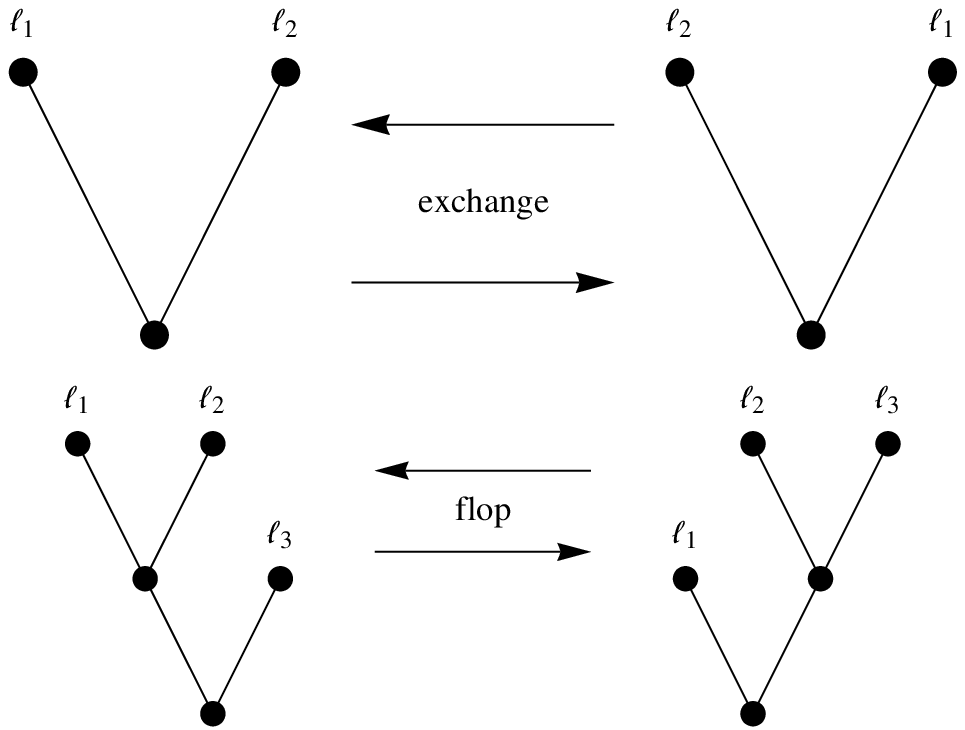}
\caption{\label{fig5}Illustration of the operations ``exchange" and ``flop" on coupling trees.}
\end{figure}

Most definitions and propositions of the sections \ref{sec:T} and \ref{sec:C} can be taken over directly
or with minor modifications. It will be appropriate to reserve the union and grafting operations, see figures \ref{fig1}
and \ref{fig3}, to binary trees, and to obtain the corresponding coupling trees by adducing a suitable permutation
of the leaves, as explained above.
Note, that the definition of ${\mathcal A}^T$ remains unchanged. We will extend the definitions
(\ref{C4a}) and (\ref{C4b}) to coupling trees by
\begin{equation}\label{A1.1}
\Delta^{\tilde{T}}=\tilde{\pi}\circ\Delta^{T}
\;,
\end{equation}
where $\tilde{T}=(T,\pi)$ and $\tilde{\pi}:{\mathcal A}^T\longrightarrow {\mathcal A}^T$
denotes the natural representation of $\pi$ by a permutation of factors of the tensor product.
Following \cite{FPJ94} we consider two operations on coupling trees, namely
\begin{eqnarray}\label{A1.2a}
\mbox{exchange:}&\quad& {\sf V}(\ell_1,\ell_2) \rightleftarrows {\sf V}(\ell_2,\ell_1)\\ \label{A1.2b}
\mbox{flop:}&\quad& {\sf V}({\sf V}(\ell_1,\ell_2),\ell_3) \rightleftarrows {\sf V}(\ell_1,{\sf V}(\ell_2,\ell_3)))
\;,
\end{eqnarray}
see figure \ref{fig5}. We have the following
\begin{proposition}\label{P1}
Let $T_1$ and $T_2$ be two coupling trees with ${\mathcal L}(T_1)={\mathcal L}(T_2)$. Then $T_1$ can be transformed
into $T_2$ by a finite sequence of exchanges and flops operating on subtrees.
\end{proposition}
We will skip the proof which is lengthy but straight forward. Note that, in the language of monomials,
the proposition says that any two monomials with the variables $\ell_1,\ldots,\ell_L$ occurring exactly once
can be transformed into each other by applying the rules of commutativity and associativity of the multiplication.\\

We will say that the coproduct $\Delta:{\mathcal A}\longrightarrow{\mathcal A}\otimes {\mathcal A}$ is
\underline{co-commutative} iff $\Delta^{{\sf V}(\ell_1,\ell_2)}=\Delta^{{\sf V}(\ell_2,\ell_1)}$ and
\underline{co-associative} iff $\Delta^{{\sf V}({\sf V}(\ell_1,\ell_2),\ell_3)}=\Delta^{{\sf V}(\ell_1,{\sf V}(\ell_2,\ell_3))}$.
Note that the coproduct defined in (\ref{M8}) is co-commutative as well as co-associative.
Obviously, $\Delta$ is co-commutative and co-associative iff $\Delta^T$ is invariant under exchanges and flops operating on
sub-trees of $T$. Together with proposition \ref{P1} we obtain:
\begin{proposition}\label{P2}
If $\Delta $ is co-commutative and co-associative and ${\mathcal L}(T_1)={\mathcal L}(T_2)$ then $\Delta^{T_1}=\Delta^{T_2}$.
Moreover, if $n_1\in{\mathcal N}(T_1)$ and $n_2\in{\mathcal N}(T_2)$ such that
${\mathcal L}(n_1)={\mathcal L}(n_2)$ then $\Delta^{n_1}=\Delta^{n_2}$.
\end{proposition}

Now let ${\mathcal L}(T_1)={\mathcal L}(T_2)$ and consider the involutive sub-algebra ${\mathcal C}_1\subset {\mathcal A}^{T_1}$
generated by the elements $\Delta^n(c),\; n\in {\mathcal N}(T_1)$ and $\Delta^{r_1}(x)$,
analogously for ${\mathcal C}_2\subset {\mathcal A}^{T_2}$, see section \ref{sec:M}. Define
\begin{equation}\label{A1.3}
CN(T_1,T_2)=\{(n_1,n_2)|n_1\in {\mathcal N}(T_1),\; n_2\in {\mathcal N}(T_2),\;{\mathcal L}(n_1)={\mathcal L}(n_2)\}
\;,
\end{equation}
and ${\mathcal C}_{12}$ as the sub-algebra of ${\mathcal C}_1\cap{\mathcal C}_2$ generated by the elements
$\Delta^{n_1}(c),\;\Delta^{r_1}(x)$ or, equivalently, by the $\Delta^{n_2}(c),\;\Delta^{r_2}(x)$,
where $(n_1,n_2)$ runs through $CN(T_1,T_2)$. Then we conclude the main result of this appendix:
\begin{theorem}\label{T3}
Let $\Delta$ be co-commutative and co-associative and $H\in{\mathcal C}_{12}$, then
$[H,K\}=0$ for all $K\in{\mathcal C}_1\cup{\mathcal C}_2$.
\end{theorem}

The scenario for superintegrability considered in \cite{BHMR04}, \cite{BB08} results as a special
case of theorem \ref{T3} in the following sense. Let $T_1$ be the ``left-homogeneous tree" and
$T_2$ the ``right-homogeneous tree" represented in figure \ref{fig2}. Then
$CN(T_1,T_2)=\{(r_1,r_2)\}$ and ${\mathcal C}_{12}$ is the algebra generated by $\Delta^{r_1}(x)=\Delta^{r_2}(x)$.
Note that in this case $T_1$ can be transformed into $T_2$ using only flops operating on sub-trees,
hence the assumption of $\Delta$ being co-commutative will be superfluous.

\section{Loop coproducts}
Recently, a framework for integrability using so-called ``loop coproducts" has been
proposed by F.~Musso \cite{M09a}, \cite{M09b} which contains different previous approaches to
integrability as special cases, namely the coalgebra approach \cite{BR98}, the linear r-matrix formulation and formulations
using Sklyanin or reflection algebras. It is, however, confined to the classical case. Nevertheless,
one may ask whether, in the classical case, the loop coproduct approach also includes
the generalization of the coalgebra approach we have given in this paper.\\
At first glance, the answer seems to be ``no", since the
corresponding derivation in \cite{M09a} of the coalgebra as a
special case utilizes the co-associativity of $\Delta$, which is not
needed in our theorem \ref{M0}. Here we neglect the differences due
to the assumption in \cite{M09a} and \cite{M09b} that the algebra
${\mathcal A}$ has a finite number of generators. A closer
inspection, however, reveals that  co-associativity is not
necessary.\\
The loop coproduct approach \cite{M09b} is based on a family of maps
$\Delta^{(k)}:{\mathcal A}\longrightarrow {\mathcal B},\;k=1,\ldots,m$
and postulates different properties of these maps
for the cases $i<k$ (or $k<i$) and $i=k$.  For comparison we have to set
${\mathcal B}={\mathcal A}^T$.
In our approach, the set of nodes is only partially ordered by the definition
$i\prec k$ iff ${\mathcal L}(i)\subset{\mathcal L}(k)$.
However, $\prec$ can be extended to a linear order $<$ such that
$i<k$ implies ${\mathcal L}(i)\subset{\mathcal L}(k)$.
For $i<k$ we have either ${\mathcal L}(i)\cap {\mathcal L}(k)=\emptyset$
or ${\mathcal L}(i)\subset{\mathcal L}(k)$. In the first case
$[\Delta^i(x),\Delta^k(y)\}=0$ for all $x\in{\mathcal A}$ due to lemma \ref{L4}.
In the second case lemma \ref{L3} implies
\begin{equation}\label{A2.1}
[\Delta^i(x),\Delta^k(y)\}=\sum_j f_j\;[\Delta^i(x_j),\Delta^i(y)\}
\end{equation}
for all $x,y\in{\mathcal A}$ and some suitable $f_j\in{\mathcal B},\;x_j\in{\mathcal A}$.
Hence, in both cases condition ($4$) of \cite{M09b} is satisfied. If $i=k$,
condition ($5$) of \cite{M09b} follows since $\Delta^i$ is a (Poisson) algebra
homomorphism in our theory.\\

We conclude that, in the case of classical mechanics and up to minor differences
in the formulations, the loop coproduct theory \cite{M09a,M09b} contains the
binary tree approach as a special case. Nevertheless, the binary tree approach has, to our opinion,
its virtues as a constructive method particularly adapted to quantum spin systems.

\end{document}